
\input phyzzx
\overfullrule=0pt

\def\Re{{\rm Re}}
\def\Im{{\rm Im}}

\rightline{COLUMBIA-UCLA/93/TEP/1}
\bigskip
\bigskip

\centerline{{\bf MOMENTUM ANALYTICITY AND FINITENESS OF THE 1-LOOP}}
\smallskip
\centerline{{\bf SUPERSTRING AMPLITUDE}
\footnote{*}{Work supported in part by NSF grants PHY-89-15286 and
DMS-92-04196.}}
\bigskip

\centerline{{\bf Eric D'Hoker}
\footnote{**} {Electronic Mail Address: DHOKER@UCLAHEP.BITNET.}}

\centerline{{\it Physics Department}}
\centerline{{\it University of California, Los Angeles}}
\centerline{{\it Los Angeles, California 90024-1547, USA}}

\bigskip

\centerline{{\bf D. H. Phong}\footnote{***}{Electronic Mail
 Address:DP@MATH.COLUMBIA.EDU}}
\centerline{{\it Mathematics Department}}
\centerline{{\it Columbia University}}
\centerline{{\it New York, N.Y. 10027, USA}}
\bigskip
\bigskip

\centerline{\bf ABSTRACT}

The Type II Superstring amplitude to 1-loop order is given by an
integral of $\vartheta$-functions over
the moduli space of tori, which diverges for
real momenta.  We construct the analytic continuation which renders this
amplitude well defined and finite,
and we find the expected poles and cuts in
the complex momentum plane.

\vfill\eject

One of the key successes of superstring theory is the observation
early on
that the Type II superstring amplitude to one loop order does not
exhibit the tachyon and dilaton divergences that occur in the bosonic
string [1].
Nowadays, all superstring amplitudes are believed to be
finite in perturbation theory [2,3].
Yet, even the nature of the most basic one-loop amplitude for
the scattering of four massless bosonic strings, such
as the graviton, dilaton and anti-symmetric tensor, has so far
not been completely elucidated (see e.g. [4]).
This amplitude is represented by an integral over positions of
vertex operators and moduli of the torus, which
is absolutely convergent only when the Mandelstam variables
$s$, $t$ and $u$ are all purely imaginary.
For real momenta, the
integral is real and infinite.

Reality of the 1-loop amplitude would
imply the vanishing of the tree-level four point function by the optical
theorem, which is in contradiction with its known non-vanishing
expression.  Both the reality and the divergence of the integral
representation are manifestations of the same illness :  the integral
representation has not been properly analytically continued in its
dependence on external momenta.
Another way to say this is that unlike in quantum
field theory, one has not properly provided an $i \epsilon$ prescription
for the string propagators [5].
This issue was addressed in the field theory limit of
superstring theory in [6,7].

In this letter, we present the crucial steps and results in the
construction of the analytic continuation of these integral representations.
Specifically, we consider the amplitude for the scattering
of four external massless bosons, including the graviton, dilaton
and anti-symmetric tensor to 1-loop in the Type II superstring.
This is the simplest non-vanishing on-shell amplitude, and
the first non-trivial quantum gravity loop amplitude which is
both finite and unitary.
We obtain explicit formulas for the singularities in the complex momentum
plane, in the form of poles in $s$, $t$ and $u$ and cuts along the
positive real axis.  The amplitude may be represented by a
double dispersion relation and the (double) spectral density is
evaluated explicitly.
Generalization to $N$-particle one loop amplitudes is technically
more involved, but straightforward.

Our starting point is the amplitude $A(k_1,\cdots,k_4)$ for the Type II
 superstring to one
loop for four external massless bosons of momenta $k_i$, $i=1\cdots,4$. The
 amplitude has a
kinematical prefactor - the same at
tree level and irrelevant to our considerations - and a single invariant
 amplitude, given as an
integral over the positions of the vertex operators and the moduli of the
 world-sheet [1,2,3]. The
world-sheet at one-loop is a torus, which can be represented as a parallelogram
 $M_{\tau}$ in the
complex plane with corners at $0, 1, \tau, 1+\tau$, and opposite sides
 identified. The parameter
$\tau=\tau_1+i\tau_2$ is the complex modulus.
Modular invariance restricts $\tau$
to the fundamental domain $\{\tau_2>0\}/PSL(2,{\bf
Z})$, which we choose to be $D=\{\tau_2>0, |\tau|>1, |\tau_1|<1/2\}$.
Let $G(z,w)$ be the Green's function on the torus defined by
$-\partial_z\partial_{\bar z}G(z,w)=2\pi\delta(z,w)-2\pi/\tau_2$. Then
$$
A(k_1,\cdots,k_4)= A(s,t,u) =
\int_D{d^2\tau\over \tau_2^2}
\int_{M_{\tau}}\prod_{i=1}^4{d^2z_i \over \tau _2}
\prod_{i<j}\exp \bigl \{ \half s_{ij} G(z_i,z_j) \bigr \}
\eqno(1)
$$
Here we have introduced standard Mandelstam variables
$s=s_{12}=s_{34}=-(k_1+k_2)^2$, $t=s_{23}=s_{14}=-(k_2+k_3)^2$,
$u=s_{13}=s_{24}=-(k_1+k_3)^2$, where all external momenta are
on mass shell $k_i^2=0$, so that $s+t+u=0$.
The Green function may be expressed in terms of $\vartheta$ functions
$$
e^{-G(z,w)}=\biggl | {\vartheta_1(z-w,\tau)\over\vartheta_1'(0,\tau)}
\biggr | ^2\exp \bigl \{ -2\pi (\Im (z-w))^2/\tau_2\bigr \}
\eqno(2)
$$
By translation invariance we may set $z_4=0$. The region of integration
separates naturally into
6 regions defined by the various orderings of
$\Im z_1, \Im z_2, \Im z_3$.
The amplitude can be split accordingly as [6]
$$
A(s,t,u) = 2A(s,t) + 2A(t,u) + 2A(u,s)
\eqno (3)
$$
with $A(s,t)$ given by the integral in (1) restricted to the region
$\Im z_1\leq \Im z_2\leq \Im z_3$ and $u=-s-t$.
The amplitude $A(s,t)$ converges absolutely
for $-2<\Re \,s,\ \Re\,t\leq 0$.
It suffices then to analytically continue $A(s,t)$ in order to
obtain a finite amplitude $A(s,t,u)$ with
manifest duality, and this is the problem we focus on.

The first step in the analytic continuation is a careful
analysis of the singularities of (1)
as $\tau$ tends to the boundary of moduli space.
For this we need the real variables $0\leq x,y\leq1$ defined by $z=x+\tau y$,
in terms of which $G(z,w)$ becomes
$$
\eqalign{
\exp (-G(z,0))  &={1\over 4\pi^2}|q|^{y^2-y}
                 |1-e^{ 2\pi ix}q^y|^2  |1-e^{-2\pi ix}q^{1-y}|^2  R(z)\cr
R(z)            &=\prod_{n=1}^{\infty}
                  |1-q^n|^{-4}
                  |1-e^{2\pi ix}q^{n+y}|^2
                  |1-e^{-2\pi ix}q^{n+1-y}|^2 \cr}
\eqno (4)
$$
Here we have set $q\equiv e^{2\pi i\tau}$. Since $|q|<e^{-\pi\sqrt 3}$ in $D$,
 $R(z)$ is a
nowhere vanishing, bounded function of both $z$ and $q$. Thus the expected
 logarithmic singularity
of the Green's function $G(z,w)$ when $z$ approaches $w$ corresponds to the
 intermediate two factors
in (4), while the term $|q|^{y^2-y}$ describes the behavior of the Green's
 function near the boundary of moduli space.
It is now convenient to introduce the variables
$ u_1=y_1,~ u_2=y_2-y_1,~ u_3=y_3-y_2,~ u_4=1-y_3 $ and
$  \alpha_i=2\pi (x_i+y_i\tau_1)$, $ i=1,2,3$ and
$\alpha _4 = 2\pi \tau _1 -\alpha _1 -\alpha _2 -\alpha _3$
and to rewrite the amplitude $A(s,t)$ as
$$
A(s,t)=\int_D{d^2\tau\over\tau_2^2}
\int_0^1 \prod_{i=1}^3{ d\alpha_i \over 2 \pi }
\int_0^1 \prod_{i=1}^4du_i
\delta(1-\sum_{i=1}^4u_i)|q|^{-(su_1u_3+tu_2u_4)}
{\cal I}(u,\alpha,q){\cal R}(u,\alpha,q)
\eqno (5)
$$
where we have defined the functions
$$
\eqalign{
{\cal I} (u,\alpha,q)=
                        &|1- e^{i \alpha _1 }|q|^{u_1 }|^{-t }
{}~|1- e^{i \alpha _2 +i\alpha _3 +i\alpha _4 } |q|^{u_2 +u_3+u_4 }|^{-t }
{}~|1- e^{i \alpha _1 +i\alpha _2 }|q|^{u_1 +u_2 }|^{s+t}\cr
                        &|1- e^{i \alpha _2 }|q|^{u_2 }|^{-s }
{}~|1- e^{i \alpha _1 +i\alpha _3 +i\alpha _4 } |q|^{u_1 +u_3+u_4 }|^{-s }
{}~|1- e^{i \alpha _1 +i\alpha _4 }|q|^{u_1 +u_4 }|^{s+t}\cr
                        &|1- e^{i \alpha _3 }|q|^{u_3 }|^{-t }
{}~|1- e^{i \alpha _1 +i\alpha _2 +i\alpha _4 }|q|^{u_1 +u_2+u_4 }|^{-t }
{}~|1- e^{i \alpha _2 +i\alpha _3 }|q|^{u_2 +u_3 }|^{s+t}\cr
                        &|1- e^{i \alpha _4 }|q|^{u_4 }|^{-s }
{}~|1- e^{i \alpha _1 +i\alpha _2 +i\alpha _3 } |q|^{u_1 +u_2+u_3 }|^{-s }
{}~|1- e^{i \alpha _3 +i\alpha _4 }|q|^{u_3 +u_4 }|^{s+t} \cr
{\cal R}(u,\alpha,q)=&
                \biggl ( {R(z_2-z_1)R(z_3) \over R(z_3-z_1)R(z_2)} \biggr
)^{-s}
                \biggl ( {R(z_3-z_2)R(z_1) \over R(z_3-z_1)R(z_2)} \biggr
)^{-t}
 \cr}
\eqno (6)
$$
A preliminary key observation is that for fixed $\tau$, the integrals
in $\alpha_i, \ u_i$ can be analytically continued to meromorphic
 functions of $s$ and
$t$ on the whole plane. In fact the condition
$|q^{u_1+u_2+u_3+u_4}|=|q|\leq e^{-\pi\sqrt 3}$ guarantees that at most 6 of
the
 12 factors in
${\cal I}(u,\alpha,q)$ can approach 0 simultaneously, so that the
following lemma and its variants will do the job :

\medskip
\noindent{\bf Lemma 1}. The integral over three complex variables $\lambda_i$,
 $i=1,2,3$ given by
$$
\int_{|\lambda_i|<1}\!\! d^2\lambda_i
|1-\lambda_1|^{-t}|1-\lambda_2|^{-s}
|1-\lambda_3|^{-t}|1-\lambda_1\lambda_2|^{s+t}
|1-\lambda_2\lambda_3|^{s+t}
|1-\lambda_1\lambda_2\lambda_3|^{-s}E(\lambda_i)
\eqno (7)
$$
where $E(\lambda_i)$ is a smooth function, can be
analytically continued
to the whole plane, with at most poles for $s$, $t$, and $u=-s-t$ at
 half-integers greater or equal to 2.
The coefficients of a pole in one variable are entire
in the other variable.
\medskip

We should note here though that these meromorphic terms without
cuts are required as part of the amplitude simply in view
of the fact that they are the ones that reproduce
correctly the large $s$ and $t$ behavior of the one-loop
amplitudes [8].
The remaining contributions with cuts have
only power law behavior.

We concentrate now on the analytic continuation of terms requiring cuts, and
 ignore meromorphic
terms which can be treated by Lemma 1.
This means that only the region with $\tau _2 \to \infty$ is relevant,
and we restrict the domain $D$
 of integration to
the simpler $\{|\tau_1|<1/2, \tau_2>1\}$. The next key observation is that it
 suffices to set ${\cal
R}(u,\alpha,q)=1$. We shall show that this case leads to branch cuts in $s$ and
 $t$ starting at 0.
In general ${\cal R}$ can be expanded up to any order $O(|q|^N)$ into a
 polynomial in $q^{n+u_i}$
with $n\geq 1$, which can be treated term by term in a similar way. The net
 effect of each term is
just to shift the beginning point of each cut to an even positive integer
 instead of 0. Physically,
the cuts  start at the lowest invariant ${\it mass}^2$ at which an intermediate
 physical string state may
be produced.

The leading case corresponds to massless intermediate states, and
 the perturbative terms
to higher mass states. Thus we need only consider the following integral
$$
A(s,t)=\int_1^{\infty}{d\tau_2\over\tau_2^2}\int_0^1
\prod_{i=1}^4 { d\alpha_i \over 2 \pi} \int_0^{\infty}
\prod_{i=1}^4du_i\delta(1-\sum_{i=1}^4u_i)
|q|^{-(su_1u_3+tu_2u_4)}{\cal I}(u,\alpha,q)
\eqno(8)
$$
up to cuts that start at non-zero positive  even integers, which can
be treated perturbatively.
The integral (8) is symmetric under $(u_1\leftrightarrow u_3, \alpha_1
\leftrightarrow \alpha_3)$ and under $(u_2 \leftrightarrow u_4,
\alpha_2 \leftrightarrow \alpha_4)$, so the integration region can be
restricted
to $u_1 \leq u_3$ and $u_2 \leq u_4$ upon
including an overall factor of 4.  The remaining integral may be split into two
regions I and II defined respectively by $(u_1+u_2)\tau_2<1$ and
 $(u_1+u_2)\tau_2>1$.

In region I, the exponential $|q|^{-(su_1u_3+tu_2u_4)}$ remains a bounded
smooth function. In view of Lemma 1, the contribution of each $\tau_2$ (and
 hence the whole
integral, since the $d\tau_2/\tau_2^2$ measure is finite) is a meromorphic
 function of $s$ and $t$ in the whole plane.

The difficult region is then II, which will generate both poles and cuts.
Here however the factor ${\cal I}(u,\alpha,q)$ simplifies considerably since
 $|q|^{u_i+u_j}<e^{-2\pi}$
whenever one index $i$ or $j$ is odd and the other even, and we can expand in
 these variables.
For $s$ and $t$ in a fixed arbitrary strip, we need keep only a finite number
of
terms. Each term can be treated in the same way as the main term, producing
cuts
 in {\it both}
$s$ and $t$ shifted to the right by an even integer. Restricting ourselves to
 the main term,
we need keep only the four factors $(1-e^{ i\alpha_i}|q|^{u_i})^{power}$,
 $i=1,\cdots,4$ in
(6). At this point it is convenient to enlarge the region II back to the full
 region I+II, since the
contribution
of I with four factors is again just a meromorphic function in the entire
plane.
The angular $\alpha_i$ dependence decouples and integrates out to a
hypergeometric function. Thus up to meromorphic functions on the whole plane
and
 higher cuts
the original amplitude $A(s,t)$ can be expressed as
$$
\eqalignno{
A(s,t) &=  \int\nolimits_1^\infty {d\tau_2 \over \tau_2^2
}\int\nolimits_0^1
\prod_{i=1}^4du_i\delta(1-\sum_{i=1}^4u_i)
|q|^{-(su_1u_3 + t u_2u_4)} F({s\over 2},{s\over 2};1,|q|^{2u_2})\cr
&\qquad\times
F({s\over 2},{s\over 2};1,|q|^{2u_4})F({t\over 2},{t\over 2};1,|q|^{2u_1})
F({t\over 2},{t\over 2};1,|q|^{2u_3})
&(9)\cr}
$$
We present the analytic continuation of (9) under two forms, each with its own
 advantages. In the
first, the cuts along the positive axis can be written out explicitly in terms
 of logarithms. An
important ingredient is the Mellin transform of hypergeometric functions, which
 we define for each
fixed integer $N$ by
$$
f_N(s,\alpha)=\int_0^1dxx^{-1-\alpha}\big [ F({s\over 2},{s\over
2};1,x)-\sum_{k=0}^{N-1} C_k(s) x^k\big ],
\qquad \quad C_k(s) ={\Gamma({s\over 2} +k)^2\over\Gamma({s\over
 2})^2\Gamma(k+1)^2}
\eqno(10)
$$
The polynomials $C_k(s)$ are the expansion coefficients of $F$
and arise naturally as the residue of the tree level closed
superstring amplitude at $t=2k$.

\medskip
\noindent {\bf Lemma 2}. The function
 $\partial^n_\alpha f_N(s,\alpha)$ is the sum of
a meromorphic function ${f_{N,n}^+}(s,\alpha)$ of $\alpha$ with poles of
order $n+1$ at $N, N+1,\cdots$, and a meromorphic function
${f_{N,n}^-}(s,\alpha)$ of $s$ with simple poles at the even
integers greater or equal to $n+2,n+4,\cdots$. The dependence on the other
 variable in each case is
entire.
\medskip

\noindent
The analytic continuation of
 $A(s,0)$ to ${\bf C}\setminus{\bf R}_+$ can now be expressed simply as
$$
A(s,0)={\pi \over 6}s^3 log(-s)  \int_0^1du~(1-u)^3u^3
f_0(s,-{1\over 2}su)^2
\eqno(11)
$$
up to a meromorphic function and higher cuts.
To obtain the analytic continuation in both $s$
 and $t$, we begin by
introducing the following expression in four variables $s,s'$, $t$, and $t'$
$$
\eqalignno{B_N(s,s';t,t')
&=- 2\pi\sum_{k_1,k_2=0}^{\infty}
C_{k_1}(t)C_{k_2}(s)
\int\int_{2k_iu_i<N,u_1+u_2<1}du_1du_2\cr
&\qquad\times\bigg [\sum_{k_3,k_4}^ {2N}
C_{k_3}(t) C_{k_4} (s)
{\Phi(-s'u_1+2k_3)-\Phi(-t'u_2+2k_4)\over s'u_1-t'u_2-2k_3+2k_4}\cr
&\qquad+\sum_{k=0}^{{N\over 2}-1}
C_k(s) \Phi(-t'u_2+2k)f_{2N}(t,{1\over 2}(-s'u_1+t'u_2-2k))\cr
 &\qquad
+\sum_{k=0}^{{N\over 2}-1}
C_k(t) \Phi(-s'u_1+2k)f_{2N}(s,{1\over 2}(s'u_1-t'u_2-2k)) \bigg ]
 &(12)\cr}
$$
where the function $\Phi$ is defined by
$$\Phi(\zeta)=(\zeta(1-u_1-u_2)+2k_1u_1+2k_2u_2)^2 log (\zeta(1-u_1-u_2)+2k
_1u_1+2k_2u_2)$$
It is readily seen that $\partial^{n}_{s'}\partial _{t'}^n B_N(s,s';t,t')$ is
holomorphic in the region $\Re\, s,\ \Re\, t <2+n$,
$s',t'\in{\bf C}\setminus{\bf R}_+$,
$\Re\,s',\Re\,t'<N$, and $N$ is large enough compared to $n$. In fact
 $B_2(s,s;t,t)$ gives precisely
the desired continuation of $A(s,t)$ to the half-plane $\Re\,s,\Re\,t<2$ up to
a
 meromorphic
function. The analytic continuation of $A(s,t)$ to an arbitrary strip can be
 obtained by
integrating (12) with respect to $s'$ and $t'$, e.g., for the strip
 $|\Re\,s|,|\Re\,t|<3$ it can be expressed as
$$
A(s,t)= A(s,0) + A(0,t) + \int_0^t dt'
\int _0^s ds' \partial_{ s'}\partial_{t'} B_3(s,s';t,t') + C_1(s,t)
\eqno(13)
$$
The additional term $C_1(s,t)$ consists of functions which are either
 meromorphic or have only
logarithmic cuts starting from $s=2$ or $t=2$, and can be worked out
explicitly.
Thus (11) and (13) give the desired exact formulas for cuts starting at 0.
Cuts corresponding to higher
intermediate mass states can be obtained successively by the same method.

The second method of obtaining the analytic continuation is
to recast expression (9) in the form of a  double dispersion
 relation
$$
A(s,t) = \int\nolimits_0^\infty d\sigma
\int \nolimits_0^\infty d\tau
{\rho_{s,t}(\sigma,\tau) \over (s-\sigma)(t-\tau)}
\eqno(14)
$$
where we have again ignored meromorphic terms. The (double) spectral density
$\rho_{s,t}(\sigma,\tau)$ is given by
$$
\eqalign{
\rho_{s,t} (\sigma,\tau) = & \int\nolimits_0^\infty\!\! d\beta_1
\int\nolimits_0^\infty \!\! d\beta_2 ~\varphi(t,\beta_1) \varphi(s,\beta_2)
\int\nolimits_0^1 \!\! du_1 \int\nolimits_0^{1-u_1} \!\! du_2 ~
(1-u_1-u_2)^2\cr
&\times \int\nolimits_{x_0}^\infty \!\! dx~(x-x_0)^2
        \theta ( u_1 \sigma -2x)
        \theta ( u_2 \tau   -2x)
        \varphi(t, {u_1\sigma \over 2}-x)
        \varphi(s, {u_2\tau  \over 2}-x  )\cr}
\eqno(15)
$$
with $x_0\equiv(u_1\beta_1+u_2\beta_2)(1-u_1-u_2)^{-1}$, and $\varphi(s,\beta)$
 is the inverse
Laplace transform of the hypergeometric function
$F({s\over 2},{s\over 2};1,x)$.
Since
$ \varphi(s,\beta)=f_0(s,\beta+i\epsilon)-f_0(s,\beta-i\epsilon)$,
the spectral density is a meromorphic function of $s$ and $t$ with poles at
even
integers.
As $\sigma,\tau \rightarrow\infty, \rho_{s,t}(\sigma,\tau)$ grows linearly, so
that the integral (15) is not convergent. We can however obtain convergence by
 subtracting a
suitable meromorphic function, a procedure analogous to (13).
The net outcome is that the double dispersion integral defines an obvious
 analytic continuation to
the full $s,t$ complex plane with a cut on the real positive $s$ and $t$
axes, with known discontinuity $\rho_{s,t} (\sigma,\tau)$. As a simple example
 we return to
the case of $t=0$. Up to a meromorphic function the double dispersion
reduces to a simple dispersion relation
with spectral density
$$
\rho_s(\sigma) =
-\int\nolimits _0 ^\infty {d \tau \over \tau} \rho _{s,0} (\sigma, \tau)=
{\sigma^3\over 24} \int\nolimits_0^1 du~u^3(1-u)^3
\bigl [ f_0(s,- {1\over 2}u\sigma)\bigr ]^2\eqno(16)
$$
which gives back the singularities obtained earlier in (11).
In view of the decomposition of $f_0$ into $f_0 ^\pm$ of
Lemma 2,
these singularities have a natural
 interpretation in terms of
underlying Feynman diagrams. The contribution of $(f_0^+)^2$ contains double
 poles
in $s$ and corresponds to the vacuum polarization diagrams in the $s$
channel.  This is the
only diagram that needs subtraction.
The contribution $2f_0^+f_0^-$
corresponds to the insertion of the triangle diagram with a single simple
pole left, and finally $(f_0^-)^2$ corresponds to the box diagram, and
no poles in $s$ occur. In summary

\noindent{\bf Theorem}.
The amplitude $A(s,t)$ can be analytically continued to
the region $(s,t)\in({\bf C}\setminus{\bf R}_+)^2$.
In this region it has poles when
$s+t$ is an integer less or equal to -2. The jumps across the cut along the
positive real axis together with the underlying poles at even positive
integers can be read off either in
terms of logarithms as in (13), or in terms of a dispersion relation
 as in (14).

\bigskip
\noindent
{\bf Acknowledgements}

We wish to thank the Aspen Center for Physics for their hospitality.

\noindent
{\bf References}

\item{[1]} M.B. Green and J.H. Schwarz, Phys. Lett. 109B (1982) 444
\item{[2]} M.B. Green, J.H. Schwarz and E. Witten, {\it
        Superstring Theory}, Cambridge University Press, 1987
\item{[3]} E. D'Hoker and D.H. Phong, Rev. Mod. Phys. 60 (1988) 917
\item{[4]} D.J. Gross, in {\it Munich High Energy Physics 1988}, Proceedings
        (1988)
\item{[5]} K. Aoki, E. D'Hoker and D.H. Phong, Nucl. Phys. B342 (1990) 149
\item{[6]} M.B. Green, J.H. Schwarz and L. Brink, Nucl. Phys. B219 (1983) 437
\item{[7]} J.L. Montag and W.I. Weisberger, Nucl. Phys. B363 (1991) 527
\item{[8]} D.J. Gross and P. Mende, Nucl. Phys. B303 (1988) 407

\vfill\eject
\end